\documentclass[conference]{IEEEtran}
\IEEEoverridecommandlockouts
\usepackage{amsmath, amssymb}
\usepackage{cite}
\usepackage{algorithmic}
\usepackage{graphicx}
\usepackage{xcolor}

\usepackage{textcomp}
\usepackage{tikz}
\usepackage{balance}
\usepackage{pgfplots}
\usepackage{pgfplotstable}
\pgfplotsset{compat=1.7}
\usepgfplotslibrary{groupplots}
\usetikzlibrary{spy}
\usepackage{nicematrix}

\ifCLASSINFOpdf
\else
\fi

\begin{document}
%
\title{Autonomous Self-Trained Channel State Prediction Method for mmWave Vehicular Communications\\

\thanks{Corresponding author: orimogunjea@run.edu.ng}}

\author{\IEEEauthorblockN{Abidemi Orimogunje\IEEEauthorrefmark{1}, Vukan Ninkovic\IEEEauthorrefmark{2}\IEEEauthorrefmark{3}, Evariste Twahirwa\IEEEauthorrefmark{1}, Gaspard Gashema\IEEEauthorrefmark{1}, Dejan Vukobratovic\IEEEauthorrefmark{2} 
\vspace{1mm}
\IEEEauthorblockA{\IEEEauthorrefmark{1}{African Center of Excellence in IoT,
University of Rwanda, Rwanda}
\IEEEauthorblockA{\IEEEauthorrefmark{2}Faculty of Technical Sciences, University of Novi Sad, Serbia}
\IEEEauthorblockA{\IEEEauthorrefmark{3}The Institute for Artificial Intelligence Research and Development of Serbia, Serbia
}}}}

\maketitle

\maketitle
\begin{abstract}
Establishing and maintaining 5G mmWave vehicular connectivity poses a significant challenge due to high user mobility that necessitates frequent triggering of beam switching procedures. Departing from reactive beam switching based on the user device channel state feedback, proactive beam switching prepares in advance for upcoming beam switching decisions by exploiting accurate channel state information (CSI) prediction. In this paper, we develop a framework for \emph{autonomous self-trained} CSI prediction for mmWave vehicular users where a base station (gNB) collects and labels a dataset that it uses for training recurrent neural network (RNN)-based CSI prediction model. The proposed framework exploits the CSI feedback from vehicular users combined with overhearing the C-V2X cooperative awareness messages (CAMs) they broadcast. We implement and evaluate the proposed framework using deepMIMO dataset generation environment and demonstrate its capability to provide accurate CSI prediction for 5G mmWave vehicular users.  CSI prediction model is trained and its capability to provide accurate CSI predictions from various input features are investigated.
\end{abstract}


%
\IEEEpeerreviewmaketitle

\vspace{7pt}
\section{Introduction}

5G millimeter wave (mmWave) technology relies on beamforming to ensure seamless and reliable connectivity between the base stations (gNB) and the user’s equipment (UE)\cite{b1,b2}. To ensure beam alignment between gNB and UE, beam management procedures exploit channel state information (CSI) between gNB and UE to perform beam selection \cite{b3,b7}. However, maintaining accurate mmWave beam alignment is a complex task even for the case of stationary devices, and it becomes significant challenge for dynamically changing conditions experienced in 5G mmWave vehicular scenarios. This calls for proactive beam management based on CSI prediction rather than traditional reactive beam management based on experienced CSI \cite{b31}. Accurate CSI prediction as an input to the mmWave beam management may not only improve the beam alignment accuracy, but also minimize the overhead in signalling used in the beam management procedures \cite{b4}. The impact of channel prediction to beam management accuracy has been part of recent studies for increasing the robustness and resilience of beam management procedures \cite{b5, b51}. 

Predicting wireless channel conditions for vehicular users is a challenging task due to highly dynamic channel behaviour \cite{b8}. Channel prediction has been addressed in a number of recent studies from different perspectives, e.g., by exploiting secondary data (e.g., aided by visual data or CSI acquired from different bands) or use of incomplete CSI data, usually paired with deep learning (DL) methods \cite{b31,b5,b8,b6,b11,b12}. Early works on 5G mmWave beamforming consider usage of historical CSI information to predict gNB beams for high-mobility vehicular users \cite{b1}. In \cite{b11}, a DL-based approach was used to predict path loss in a vehicular communication environment using datasets created by a hybrid combination of network simulator 3 (NS3) and Simulator of Urban Mobility (SUMO). The prediction of mmWave channel in vehicle-to-vehicle (V2V) scenario using autoregressive techniques is carried out in \cite{b8}, exploiting the CSI dataset collected from vehicles moving towards each other. The proposed mobility-based channel prediction module uses the mobility parameters of the vehicles to perform channel prediction. In \cite{b12}, the authors applied recurrent neural networks (RNN) to predict CSI in a time-varying channel, reducing the amount of pilot symbols required for efficient channel estimation.

In this paper, we focus on developing a CSI prediction module for 5G mmWave vehicular users as a support for mmWave beam management module at gNB. The proposed framework aims for a fully autonomous module that continuously collects and labels acquired data, generates and maintains a training data set, and triggers the model training to generate or update the CSI prediction module. The collected data set consists of: i) CSI data set collected from 5G mmWave vehicular UEs, e.g., by exploiting demodulation reference symbols (DM-RS) in the cell area, and ii) location, speed and acceleration data set collected by overhearing V2V cooperative awareness messages (CAM) \cite{b12-1}. Our assumption is that the gNB is equipped with Rel. 14 C-V2X receiver overhearing the shared $5.9$ GHz V2V channel and decoding all CAM messages exchanged among vehicles in the surrounding of gNB. Time synchronised data from both sources (CSI and CAM-based) are locally stored at gNB for training the prediction model, making the proposed framework operating essentially in autonomous and self-training mode by simultaneously logging time-synchronised input data and their corresponding labels. In this paper, we use a slight modification of DeepMIMO software \cite{b13} to recreate the above environment and generate realistic data collected at gNB. From acquired data, we train an RNN architecture with long short-term memory (LSTM) layers as a CSI prediction module. Extensive numerical evaluation demonstrates that the proposed framework is feasible and that the RNN-based CSI prediction model provides accurate CSI prediction in a realistic DeepMIMO-generated 5G mmWave vehicular environment.    

The rest of the paper is organised as follows. Sec. II presents the system model. In Sec. III, the proposed CSI prediction framework for 5G NR mmWave vehicular users is presented. Sec. IV presents simulation setup based on DeepMIMO modification, CSI prediction model training and performance evaluation. The paper is concluded in Sec. V. 

\section{System Model}

In this section, we set the 5G mmWave vehicular communications system model under consideration.

\begin{figure}
  \centering
\includegraphics[width=0.5\textwidth]{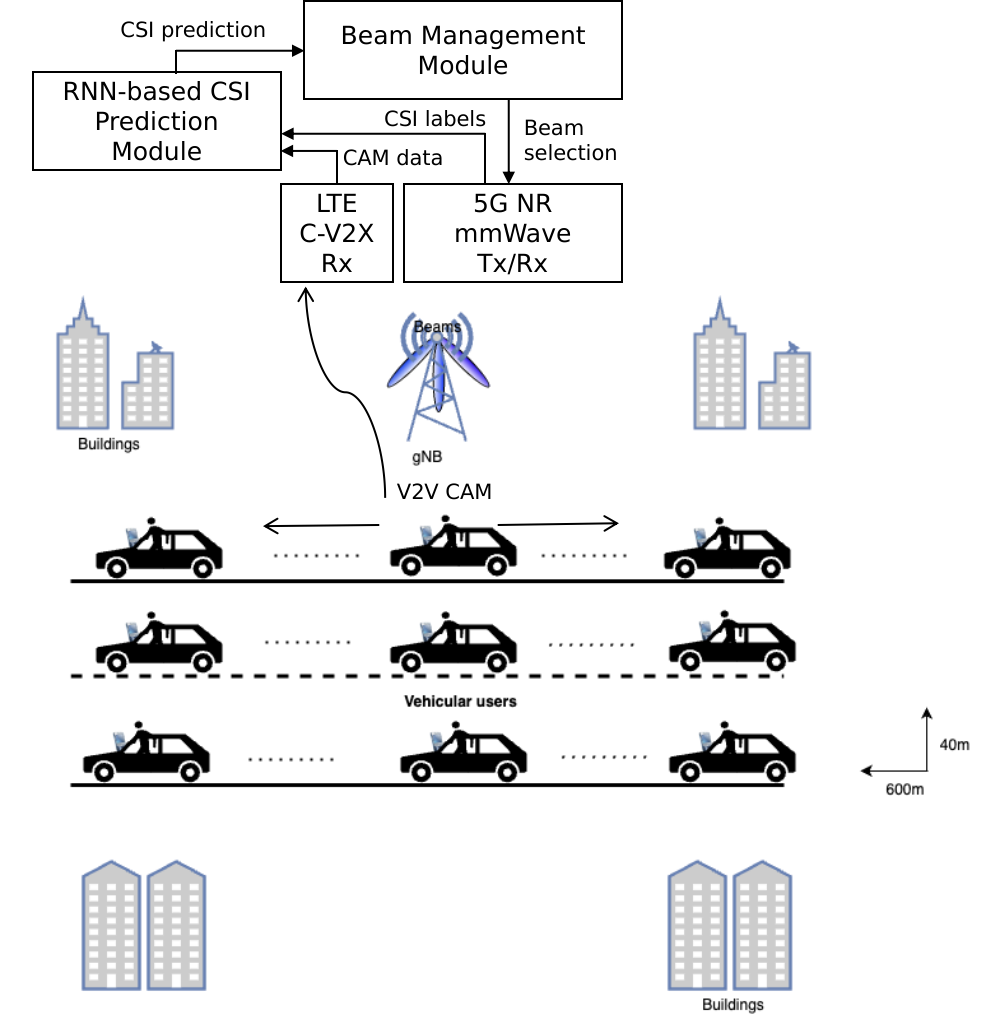}
  \caption{System model for vehicular users in 5G NR mmWave environment}
  \label{Fig_1}
\end{figure}

\textbf{5G mmWave Base Station Model:} We consider a 5G NR mmWave gNB deployed in an urban environment covering an area of a single busy road/street. gNB operates in mmWave band, i.e., frequency range 2 (FR2) at the carrier frequency $f_c$. Its antenna panel is a 2D uniform linear array consisting of $M=M_x \times M_y$ antennas at a distance of $d=\lambda/2$ from each other in both dimensions, where $\lambda$ is the wavelength corresponding to $f_c$. gNB serves vehicular UEs (described later) using 5G mmWave beamforming in the downlink channel of bandwidth $B$ containing $K$ OFDM subcarriers, where beam management decisions are made by a downlink Beam Management Module (DL-BMM). We assume DL-BMM exploits periodic CSI prediction inputs it receives from CSI Prediction Module (CSI-PM) for every attached vehicular UE, where periodicity of CSI prediction delivery can be configured for each UE. The focus of our work is CSI-PM that generates per-UE CSI predictions. This module collects data from two radio interfaces: 1) From 5G NR FR2 interface, the CSI feedback obtained using channel reference symbols (e.g., CSI feedback obtained through DM-RS symbols) is delivered to CSI-PM module along with the exact acquisition timing (e.g., frame and subframe index), and 2) From additionally integrated LTE C-V2X receiver, gNB overhears cooperative awareness messages (CAMs) assumed to be broadcasted within the $5.9$~GHz band periodically with period $\Delta \tau$ (typically, $\Delta \tau = 100$~ms) by all vehicular UEs in the cell from which it extracts exact timing, location, speed and acceleration of each UE. The two data streams are collected, aligned in time and stored in a dataset that is used for training the CSI prediction model: a core component of CSI-PM. Note that the process is essentially autonomous, i.e., the data collection and labeling process can be performed automatically where vehicular UE features such as location, speed, acceleration are associated with the corresponding CSI label taking into account their time alignment. Finally, we note that both CSI-PM and BMM modules can be implemented as eXtended Applications (xApps) within O-RAN Near-Real-Time Radio Interface Controller (RIC) \cite{b13-1}.   

\textbf{Channel and CSI Model:} We consider a ray-tracing based channel model developed for a given environment (3D model of environment and location of gNB and UEs) using deepMIMO simulator \cite{b13}. If the user is equipped with $N$ antennas, the channel matrix $\mathbf{H}^{b,u}_{k} \in \mathbb{C}^{M \times N}$ is constructed between a gNB $b$ and a user $u$ at subcarrier $k$. Taking into account $L$ strongest ray-tracing paths, and for simplicity, assuming that the number of receive antennas $N=1$, the vector of channel coefficients $\mathbf{h}^{b,u}_{k} \in \mathbb{C}^{M \times 1}$ between a gNB $b$ and a user $u$ at subcarrier $k$ is given as:
\begin{equation}
\mathbf{h}_k^{(b,u)} = \sum_{l=1}^{L} \sqrt{\frac{\rho_l}{K}} e^{j\left(\vartheta_l + \frac{2\pi k}{K} \tau_l B \right)} \mathbf{a}\left(\varphi_{\text{az}}^{(b,u)}, \varphi_{\text{el}}^{(b,u)}\right)
\end{equation}
where: $\rho_l$ and $\tau_l$ is the path gain and propagation delay ($l$ is the channel path index), $B$ is the bandwidth,
$\vartheta_l$ is the phase of the signal path,
$\mathbf{a}\left(\varphi_{\text{az}}^{(b,u)}, \varphi_{\text{el}}^{(b,u)}\right)$ is the array vector of the gNB, expressed in Kronecker product form as:
\begin{align}
\mathbf{a}(\varphi_{\text{az}}^{(b,u)}, \varphi_{\text{el}}^{(b,u)}) & =  \mathbf{a}_x(\varphi_{\text{az}}^{(b,u)}, \varphi_{\text{el}}^{(b,u)}) \otimes \nonumber \\
& \otimes \mathbf{a}_y(\varphi_{\text{az}}^{(b,u)}, \varphi_{\text{el}}^{(b,u)}) \otimes \mathbf{a}_z (\varphi_{\text{el}}^{(b,u)}),
\end{align}
where $\mathbf{a}_x(\cdot)$, $\mathbf{a}_y(\cdot)$ and $\mathbf{a}_z(\cdot)$ represent the BS array response vector in $x$, $y$ and $z$ direction, respectively (see \cite{b13} for more details). Although in practice, a quantised version of CSI is fed back, for simplicity, we assume that gNB receives and collects full CSI information in the form of a matrix $\mathbf{H}^{b,u} \in \mathbb{C}^{M \times K}$ that collects channel gains across all $K$ subcarriers.





\textbf{Vehicular Users Model:} We consider a fleet of vehicular users moving along the road in both directions. Each vehicle is equipped with 5G NR mmWave (FR2) and LTE C-V2X 5.9 GHz radios for V2I and V2V communications, respectively. gNB maintains connectivity with vehicular UEs through beam management process implemented in the BMM module. Vehicular UEs broadcast CAM messages with periodicity $\Delta \tau$. 

We adopt a simple mobility model for vehicular mobility where a vehicle acceleration changes according to a finite-state Markov Chain (FSMC) model with $2S+1$ states. The set of states contains a set of equidistant acceleration values in the interval $[-a, a]$ with step $\Delta a=a/S$. Starting from a random location at one of the street ends, the vehicle location, speed and acceleration is initialised, where speed is randomly uniformly sampled from an interval $[v_{min}, v_{max}]$ and acceleration is randomly set to one of $2S+1$ FSMC states. From this information, and assuming constant acceleration during the period of duration $\Delta \tau$, a new location and speed after $\Delta \tau$ period is calculated and assumed to be broadcasted (and overheard by gNB) through a CAM message. At that time instant, FSMC moves from acceleration state $a$ to each of the two neighbouring acceleration states with probability $p$, or remains in the same acceleration state with probability $1-2p$ (except the edge states where it remains with probability $1-p$). By controlling $p$, we control the dynamics of acceleration and deceleration. Although the considered FSMC-based mobility model is simple and does not capture many features of vehicular mobility, the model captures basic movement patterns of vehicles that is easily integrated into the DeepMIMO simulator to collect a joint mobility/CSI dataset.   

\section{CSI Prediction Module for 5G NR mmWave Vehicular Users}

In this section, we further elaborate a framework for data collection, training and inference for CSI prediction implemented as a stand alone CSI-PM module. The CSI-PM module performs several main functions, as detailed below.

\textbf{Data collection and labelling:} For each vehicular UE connected to the gNB, the CSI-PM module collects time-stamped CSI information received from 5G NR mmWave radio and time-stamped content of CAM messages containing UE location, speed and acceleration. Time aligned data (of periodicity $\Delta \tau$) aggregated across all UEs is stored in a training data set where CAM data represents features and CSI data represents labels. The dataset generation process (including labelling) is fully automated.

\textbf{CSI-PM model architecture and training:} The core of the CSI-PM module is the prediction model. We train the prediction model on the collected dataset. The proposed model architecture adheres to conventional methodologies for training multivariate time series models, employing recurrent neural networks (RNNs) as our baseline model. Specifically, we utilize an RNN architecture with two long short-term memory (LSTM) layers (each with 10 hidden states), regardless of the number of input features. This decision was motivated by our pursuit of a comprehensive understanding of the model's robustness and expressive capabilities. Additionally, maintaining a consistent architecture across varied input features offers the advantage of simplifying feature management and model interpretation. While we recognize the potential trade-offs in model performance due to this uniform approach, practical considerations, such as resource constraints and the desire for a unified framework, influenced our decision-making. By adhering to this approach, we aim to explore the generalizability and adaptability of our model across diverse datasets and application scenarios. 

In our RNN architecture, we incorporate a fully-connected (FC) layer as the output layer alongside the two LSTM layers. The FC layer comprises $2M$ neurons, with each neuron dedicated to processing specific features, encompassing $M$ real and $M$ imaginary CSI values (equal to the number of predicted channel coefficients). By embedding this FC layer at the output stage of the RNN, our aim is to refine the model's predictive prowess by adeptly capturing nuanced data patterns. This augmentation empowers the algorithm to generate precise predictions grounded in a comprehensive analysis of the input features. Consequently, the synergy between the LSTM layers and the FC output layer strengthens the model's resilience and ensures accurate predictions, thereby enhancing its efficacy in handling multivariate time series data. The model is optimized through mean squared error (MSE) loss function (between estimated and true CSI values), on a batch-by-batch basis, using the Stochastic Gradient Descent (SGD) algorithm with the ADAM optimizer \cite{b16}.

\textbf{CSI-PM model inference:} Once the model is trained, it is deployed in the CSI-PM module for inference process. Based on inputs obtained from each UE after each received CAM message, the CSI-PM module generates CSI predictions and sends them to the BMM module to support beam management decisions. We note that a retraining process based on the newly received and stored data can be trigerred periodically to update the CSI-PM model.

\section{Performance Evaluation}

\subsection{Simulation Setup for 5G NR mmWave Environment}

The scenario considered in this paper (Figure \ref{Fig_1}) is implemented in DeepMIMO by performing minor adaptations to the simulator by including simulation of mobile vehicular UEs according to a simple FSMC mobility model. For the simulation model, we consider the main street called User grid 1 (UG1) as part of the first scenario (Scenario 1) in the DeepMIMO simulator. The street is about $600$~m long and $40$~m wide. More precisely, it is represented as a grid of possible discretised user positions with precisely $2751$ rows and $181$ columns and a distance of $0.2$~m between neighboring grid points in both dimensions. A single gNB deployed in the street at one of the central street locations (BS3 in DeepMIMO Scenario 1) is activated. gNB parameters and their values adopted in the simulation are presented in Table 1. 

\begin{table}[htbp]
    \caption{DeepMIMO gNB Parameters}
    \label{tab:example}
    \centering
    \begin{tabular}{|c|c|}
        \hline
        DeepMIMO parameter & Value \\
        \hline
        Number of gNB antennas $M$ & $16 (4 \times 4)$ \\
        Antenna spacing $d$ & $\lambda/2$  \\
        System bandwidth $B$ & $100$~MHz \\
        No. of OFDM subcarriers $K$ & 240 \\
        Carrier Frequency $f_c$ & $28$~GHz \\
        Number of Ray-Tracing Paths $l$ & $5$ \\
        \hline
    \end{tabular}
\end{table}
We adapted the DeepMIMO simulator for the purpose of this work by introducing and tracking the movement of vehicles along the main street. Users are generated at a random column and follows the same column along the street moving according to the FSMC mobility model adopted in Sec. II. Vehicle speed is limited to the interval $[v_{min}, v_{max}]=[30,50]~km/h$, and its acceleration is generated from FSMC model with $2S+1=5$ states where $[-a, a]=[-1,1]$~m/s and $S=2$. The FSMC dynamics is governed by probability $p=0.2$ of transitioning to neighbouring states. For simplicity, the Doppler effect on CSI values is neglected (it will be included in our future study). A large number of simulated vehicle locations sampled every $\Delta \tau$ and their corresponding CSI values are generated by DeepMIMO simulator as a realistic data set assumed to be collected by CSI-PM module for training the CSI prediction model.  

\begin{table*}[tbhp]
\caption{Model MSE Regarding Input Feature Selection.}
\begin{center}
\begin{tabular}{|c|c|c|c|c|c|c|c|c|c|c|}
\hline
MSE/Feature&Acc. & Speed &Pos. & Acc.+Speed & Acc.+Pos.& Speed+Pos. & Acc.+Speed+Pos. &CSI\textsubscript{1}&CSI\textsubscript{1}+Pos.&CSI\textsubscript{1}+CSI\textsubscript{2}+Pos.\\
\hline
Data Set 250&1.13e-07	
            &4.48-07	
            &1.52e-08	
            &2.66e-08	
            &8.62e-08	
            &1.46e-07	
            &5.98e-07	
            &5.03e-08	
            &5.66e-10	
            &1.04e-10\\	
\hline
Data Set 500&2.40e-07	
            &2.35e-07	
            &1.84e-08	
            &2.15e-07	
            &3.96e-08	
            &1.10e-06	
            &2.44e-07
            &6.72e-08
            &1.188-10
            &4.87e-11\\	
\hline
Data Set 750&1.92e-07	
            &1.42e-7	
            &1.97e-08	
            &2.57e-7
            &3.67e-8	
            &2.26e-7	
            &1.98e-7
            &5.43e-08
            &1.74e-10
            &4.42e-11\\
\hline
\end{tabular}
\label{table_sota}
\end{center}
\end{table*}

        


\subsection{CSI-PM Model Training Procedure}
\label{setup}
The proposed approach is evaluated across three distinct datasets, each constructed according to the methodology presented in Sec. II and detailed in Sec. IV-A. Their primary differentiation is the maximum row distance between the vehicle location and the gNB. Specifically, the row distance is set to $250$, $500$, and $750$ (or $50$, $100$, $150$~m) respectively. The aim of the three scenarios is to reveal if the quality of CSI prediction depends on the maximum range between the gNB and the UEs, where the maximum distance of $150$~m corresponds to a typical range of V2V messages (that need to be captured by the gNB). The three datasets contain $8860$, $17733$, and $27805$ instances, respectively. In all simulations, 70 \%\ of the data is allocated for training purposes, with the remaining data used for testing. Each instance encompasses information from a single position, consisting of the three feature inputs: the vehicle speed, acceleration, and position, labelled with $M=16$ complex values representing the CSI between $16$ antennas at the base station and the single antenna at the vehicle. Throughout the conducted experiments, the objective was predicting the vehicular UE CSI at its subsequent positions (separated by $\Delta \tau$ in the time domain) using inputs and CSI values gathered from the preceding and the current position. The input data is first preprocessed in such a way that the complex CSI values are separated into real and imaginary parts, and the two vectors are concatenated into a single real CSI vector of length $2M$. Vehicle mobility features (speed, acceleration, and position) are used in at most $10$ previous vehicle positions. The proposed RNN model is trained using a learning rate $\alpha=0.00008$, $\beta_1=0.9$ and $\beta_2=0.999$, while the batch size is set to 64.

\subsection{Numerical Results}
\label{NR}
Focusing on CSI prediction mean square error (MSE), we examined the performance of CSI-PM module across various input features, consolidating the obtained results in Table \ref{table_sota}. By systematically evaluating prediction MSE performance vs selected subset of input features, we gained insights into the trade-offs between feature complexity and performance. The performances showcased in Table \ref{table_sota} offer interesting insights. While datasets $500$ and $750$ pose challenges in CSI prediction due to larger UE distance and consequently lesser CSI resolution, they offer a larger volume of data compared to dataset $250$. Consequently, the CSI prediction trends observed in the behavior of results across these three datasets remains similar. More precisely, when considering individual features, position (i.e., 10 previous positions, as explained in Section \ref{setup}) stands out a low MSE (Table \ref{table_sota}), indicating its significant informational value. Despite this promising performance, the average prediction MSE values per dataset are $1.86512e-6$, $1.352e-6$, and $1.1e-6$ for datasets $250$, $500$, and $750$, respectively. Incorporating additional input features in the form of CSI values captured in one or two previous positions results in notable performance enhancement, as seen in Table \ref{table_sota}. 

\begin{figure}[t]
	\begin{tikzpicture}[spy using outlines=
{rectangle, red, magnification=3.2, line width=2, connect spies}]
  	\begin{axis}[width=0.85\columnwidth, height=6.5cm, 
	legend style={at={(0.2,0.9)}, anchor=north,font=\scriptsize, legend style={nodes={scale=0.75, transform shape}}},
   	legend cell align={left},
	legend columns=1,   	 
   	x tick label style={/pgf/number format/.cd,fixed,
   	 precision=1, /tikz/.cd},
   	y tick label style={/pgf/number format/.cd,fixed, precision=1, /tikz/.cd},
   	xlabel={Re},
   	ylabel={Im},
   	label style={font=\footnotesize},
   	grid=major,   	
   	xmin =-0.000008, xmax = 0.000006,
   	ymin=-0.000015, ymax=0.00001,
   	line width=0.8pt,
   	tick label style={font=\footnotesize},
    ylabel near ticks, yticklabel pos=right]
    \addplot[blue, only marks, mark=square] 
   	table [x={x}, y={y}] {./Picture.png/tikz/y_coor_250_real}; 
   	\addlegendentry{True CSI - 250}  \addplot[blue, only marks, mark=star] 
   	table [x={x}, y={y}] {./Picture.png/tikz/y_coor_250_pred}; 
   	\addlegendentry{Predicted CSI - 250 (Input: Position)} 
    \addplot[red, only marks, mark=triangle] 
   	table [x={x}, y={y}] {./Picture.png/tikz/CSI1_CSI2_y_coor_250_real}; 
   	\addlegendentry{True CSI - 250}  \addplot[red, only marks, mark=o] 
   	table [x={x}, y={y}] {./Picture.png/tikz/CSI1_CSI2_y_coor_250_pred}; 
   	\addlegendentry{Predicted CSI - 250 (Input: CSI\textsubscript{1}+CSI\textsubscript{2}+Pos.)}
    \coordinate (spypoint) at (axis cs:0.0,0.0000003);
\coordinate (spyviewer) at (axis cs:-0.000005, -0.0000105);
\spy[width=4.5cm,height=2cm] on (spypoint) in node [fill=white] at (spyviewer);
  	\end{axis}
	\end{tikzpicture}
	\vspace*{-3.9mm}
	\caption{Comparison of true CSI values and predicted CSI values (single instance of 16 complex values) for dataset 250 using various input features}
	\label{Fig_250}
\end{figure}
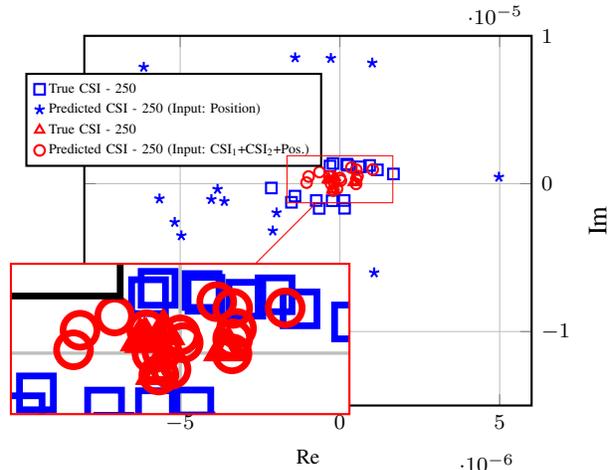
\begin{figure}[t]
	\begin{tikzpicture}[spy using outlines=
{rectangle, red, magnification=3.2, line width=2, connect spies}]
  	\begin{axis}[width=0.85\columnwidth, height=6.5cm, 
	legend style={at={(0.2,0.9)}, anchor=north,font=\scriptsize, legend style={nodes={scale=0.75, transform shape}}},
   	legend cell align={left},
	legend columns=1,   	 
   	x tick label style={/pgf/number format/.cd,fixed,
   	 precision=1, /tikz/.cd},
   	y tick label style={/pgf/number format/.cd,fixed, precision=1, /tikz/.cd},
   	xlabel={Re},
   	ylabel={Im},
   	label style={font=\footnotesize},
   	grid=major,   	
   	xmin =-0.00002, xmax = 0.00004,
   	ymin=-0.00001, ymax=0.00002,
   	line width=0.8pt,
   	tick label style={font=\footnotesize},
    ylabel near ticks, yticklabel pos=right]
    \addplot[blue, only marks, mark=square] 
   	table [x={x}, y={y}] {./Picture.png/tikz/y_coor_500_real}; 
   	\addlegendentry{True CSI - 500}  \addplot[blue, only marks, mark=star] 
   	table [x={x}, y={y}] {./Picture.png/tikz/y_coor_500_pred}; 
   	\addlegendentry{Predicted CSI - 500 (Input: Position)} 
    \addplot[red, only marks, mark=triangle] 
   	table [x={x}, y={y}] {./Picture.png/tikz/CSI1_CSI2_y_coor_500_real}; 
   	\addlegendentry{True CSI - 500}  \addplot[red, only marks, mark=o] 
   	table [x={x}, y={y}] {./Picture.png/tikz/CSI1_CSI2_y_coor_500_pred}; 
   	\addlegendentry{Predicted CSI - 500 (Input: CSI\textsubscript{1}+CSI\textsubscript{2}+Pos.)}
    \coordinate (spypoint) at (axis cs:0.0,0.0000003);
\coordinate (spyviewer) at (axis cs:0.00003, 0.00000105);
\spy[width=1cm,height=2cm] on (spypoint) in node [fill=white] at (spyviewer);
  	\end{axis}
	\end{tikzpicture}
	\vspace*{-3.9mm}
	\caption{Comparison of true CSI values and predicted CSI values (single instance of 16 complex values) for dataset 500 using various input features}
	\label{Fig_500}
\end{figure}
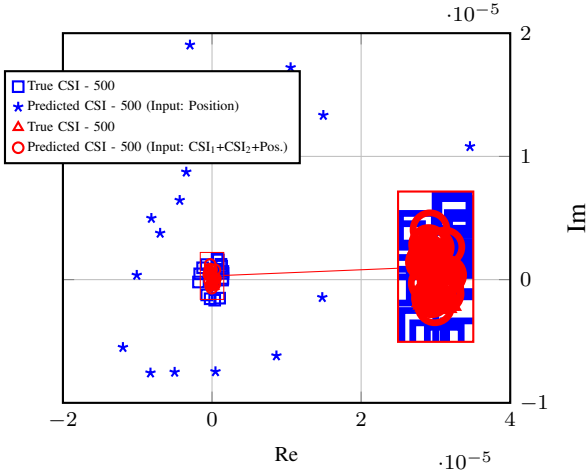

To visualize the prediction of a single instance, we utilized the proposed RNN-based model to generate the CSI predictions. Considering that the instances are complex values, we determined the closest instance from the actual test set by comparing their amplitudes using MSE and mean absolute error (MAE) metrics. It is interesting that both metrics return the same instance, within one dataset. From Fig. \ref{Fig_250}, it is evident that for dataset $250$, using the previous 10 positions as an input feature does not yield accurate estimates. This is indicated by the noticeable difference between the true CSI (blue squares) and the predicted CSI (blue stars). However, introducing CSI from two previous positions as additional input features, alongside the previous 10 positions, significantly improves performance. This improvement is evident in the red magnifying window, where the red circles (predicted CSI values) and triangles (real CSI values) nearly overlap, signifying a substantial reduction in error by several orders of magnitude. A similar conclusion can be drawn for dataset $500$ (Fig. \ref{Fig_500}), except that the introduction of previous CSI values further enhances performance, as expected from the MSE values in Table \ref{table_sota}. 

\section{Conclusion}
Accurate CSI prediction could significantly improve the performance of beam management procedures in 5G NR mmWave networks. In this work, we explore a framework for autonomous and self-trained CSI prediction implemented as a stand-alone CSI-PM module. The module collects and labels data by overhearing vehicular C-V2X CAM messages, extracts features about each vehicular UE, and label them with the corresponding CSI feedback. The obtained dataset is used to train fully connected LSTM network on the data produced by suitable adaptation of the DeepMIMO simulator. The CSI-PM model is trained and its capability to provide accurate CSI predictions from various input features is investigated. Further work will target further development and integration of CSI-PM and BMM modules in DeepMIMO and their subsequent development as xApps in an open-source O-RAN environment.

\vspace{7pt}
\section*{Acknowledgment}
This work was jointly supported by the African Center of Excellence in Internet of Things (ACEIoT) from College of Science and Technology, University of Rwanda, and The Regional Scholarship and Innovation Fund (RSIF). In addition, this work received funding from the Horizon 2020 research and
innovation staff exchange grant agreement No 10108638.




%

\end{document}